\documentclass[conference]{IEEEtran}
\IEEEoverridecommandlockouts
\usepackage{cite}
\usepackage{amsmath,amssymb,amsfonts}
\usepackage{algorithmic}
\usepackage{graphicx}
\usepackage{textcomp}
\usepackage[table,xcdraw]{xcolor}
\usepackage{multirow}
\usepackage{graphicx}
\usepackage{booktabs}

\def\BibTeX{{\rm B\kern-.05em{\sc i\kern-.025em b}\kern-.08em
    T\kern-.1667em\lower.7ex\hbox{E}\kern-.125emX}}
\begin{document}

\title{A guideline proposal for minimizing cybersickness in VR-based serious games and applications\\

{\footnotesize \textsuperscript{}}
\thanks{}
}

\author{\IEEEauthorblockN{1\textsuperscript{st} Thiago Porcino, 2\textsuperscript{nd} Derek Reilly}
\IEEEauthorblockA{\textit{Faculty of Computer Science} \\
\textit{Dalhousie University}\\
Halifax, Canada \\
thiago@dal.ca\\reilly@cs.dal.ca}
\and
\IEEEauthorblockN{3\textsuperscript{rd} Esteban Clua, 4\textsuperscript{th} Daniela Trevisan}
\IEEEauthorblockA{\textit{Institute of Computing} \\
\textit{Universidade Federal Fluminense}\\
Niteroi, Brazil \\
esteban@ic.uff.br\\daniela@ic.uff.br}

}
\maketitle

\begin{abstract}
Head-mounted displays (HMDs) are popular immersive tools in general, not limited to entertainment but also for education, military, and serious games for health. While these displays have strong popularity, they still have user experience issues, triggering possible symptoms of discomfort to users. This condition is known as cybersickness (CS) and is one of the most popular research topics tied to virtual reality (VR) issues. We first present the main strategies focused on minimizing cybersickness problems in virtual reality. Following this, we propose a guideline framework based on CS causes such as locomotion, acceleration, the field of view, depth of field, degree of freedom, exposition use time, latency-lag, static rest frame, and camera rotation.
Additionally,  serious games applications and broader categories of games can also adopt it. Additionally, we categorized the imminent challenges for CS minimization into four different items. Conclusively, this work contributes as a consulting reference to enable VR developers and designers to optimize their VR users' experience and VR serious games.

\end{abstract}

\begin{IEEEkeywords}
head-mounted displays, virtual reality, cybersickness, VR-based serious games, challenges
\end{IEEEkeywords}

\section{Introduction}


Games and Virtual Reality may be efficient environments for teaching and training in numerous fields, such as education, military, and health.
In particular, dedicated applications for health retain the purpose of training health professionals and as tools to minimize health disorders. For example, serious games can be developed for rehabilitation processes and therapies \cite{deponti2009droidglove,min2022effectiveness}. Besides, with the immersive contribution of popular XR (extended reality) supplies, such as head-mounted displays (HMDs) and the popularity of immersive environments over the last years, serious games that can employ immersive virtual conditions and gaming mechanics maintain a promising future, supported by the market of vast hardware availability for immersive games \cite{checa2020review}. 



HMDs devices are used for various purposes in the industry such as in games that focus on health \cite{carrion2019developing} and serious games \cite{checa2020review}. However, VR consumers are often complaining about frequent manifestations of discomfort in these devices \cite{kolasinski1995simulator}. This situation is known as cybersickness (CS) and is a condition that has been intrinsic in HMDs. CS can also trigger a scope of symptoms of discomfort in users.





The discomfort effect can negatively impact the VR treatment and, in some cases, even present dangerous situations for the current patient. For example, while a VR-based game for upper limb rehabilitation \cite{ferreira2020adaptive} can be used as an excellent therapy tool, it also can produce discomfort in patients immersed in a virtual environment hearing HMDs. In addition, CS symptoms can vary across individuals; some people are more susceptible than others. In summary, we start from the principle that CS symptoms are a side effect of any immersive application (including immersive serious games) that does not follow protocols to minimize the CS.

In this work, we propose a guideline to improve the user experience and minimize cybersickness symptoms. Additionally, we categorized the main challenges in this context. To build this guideline, we focus on immersive environments, including immersive serious games for health-based treatments, VR entertainment games, and VR applications. Furthermore, we highlight some recent immersive serious games that focus on rehabilitation but do not apply any strategy to produce a better user experience related to minimizing cybersickness in their application. We believe our guideline can improve patients' experience using immersive serious games and head-mounted displays. More importantly, we look forward to answering the following research questions: 
\begin{itemize}
    \item Which are the main strategies reported in the literature to minimize CS in VR serious games?
    \item Which are the main challenges for mitigating CS in VR environments? 
\end{itemize}

This paper is organized as follows: Section 2 clarifies the literature review presenting the necessary background knowledge to understand the types of VR-related illnesses. Section 3 explains the related works highlighting the importance of using strategies to avoid CS in VR-based serious games for rehabilitation. Section 4 describes our proposed guideline to mitigate discomfort in VR-based environments and serious games, presenting the strategies to overcome CS. Section 5 points out the main challenges to minimizing CS in head-mounted displays. Section 6 presents a discussion of this work's findings. Finally, section 7 describes the conclusion and future work.

\section{CS Background}

This section presents a fundamental understanding of motion perception concerning motion sickness's (MS) primary distinctions and subcategories. 
MS manifests itself because of the information divergence emitted by the human sensory system. This occurs when conflicts between the sensory organs define an individual's orientation and spatial positioning. MS is defined as the discomfort felt during a forced visual movement without equivalent body movement, such as those present in the airplane, boats, or land vehicles locomotions \cite{kemeny2020getting}. 

This discomfort also occurs in virtual environments and is called visually induced motion sickness (VIMS). Merhi et al. \cite{merhi2007motion} defined the event of VIMS during experiments with video games as a game disease (gaming sickness). Moreover, in VR, articles usually label VIMS that occurs in VR as CyberSickness (CS) \cite{mccauley1992cybersickness}. In contrast, VIMS that occurs during flight or drive simulators is often called simulator sickness \cite{brooks2010simulator}. Overall, MS can be split into two subcategories \cite{kemeny2020getting}: transportation sickness, which is tied to the real world, and simulator sickness, which is associated with the virtual world and includes cybersickness (CS). 



Cybersickness symptoms, in turn, are comparable with MS symptoms occurring in the real world, such as nausea, vertigo, dizziness, and upset stomach \cite{howarth1997occurrence}. CS Symptoms occur mainly with VR Head-Mounted Displays (HMDs) \cite{rebenitsch2015cybersickness}.
 \cite{kolasinski1995simulator} described more than 40 possible VIMS causes. These factors were grouped into simulator, task, and individual factors.
 \cite{renkewitz2007perceptual} expanded Kolasinski's work by tabling and dividing potential factors for CS manifestations. Rebenitsh \cite{renkewitz2007perceptual} stated that many factors and configurations related to discomfort are still unknown. 
 

\section{Related Works}
CS attenuation for different users and games is not trivial. The first problem is the lack of a unique variable for discomfort levels. VR users may experience multiple symptoms. Besides, serious games focused on treatment use VR environments and equipment (such as HMDs) to improve user experience and patient engagement in therapy. 

However, the lack of CS mitigation strategies in VR-based serious games may impair therapy results, generating side effects on their patients. For example, the CS symptoms generated in a VR-based game for therapy could be a negative side effect in a VR game without CS minimization strategies. For this reason, in this section, we separated related works considering works focused on CS minimization and VR-based serious games for rehabilitation.

\subsection{CS minimization in VR}

Several studies have been conducted using deep learning models, such as convolutional neural networks (CNNs) and recurrent neural networks (RNNs).  \cite{kim2019deep}, proposed a deep learning architecture to estimate the cognitive state using brain signals and how they are related to CS levels. Their approach is based on deep learning models, such as long short-term memory (LSTM), RNN, and CNN ~\cite{lawrence1997face, graves2013speech, sak2014long}). The models learn the individual characteristics of the participants that lead to the manifestation of CS symptoms when playing a VR game.

In a previous work \cite{porcino2022identifying}, we propose an approach to identifying causes of cybersickness in different virtual reality games using head-mounted displays. We used symbolic classifiers to analyze the causes of cybersickness during the gameplay experience. Once the cause is identified, game designers can select an adequate strategy to mitigate the impacts of CS.

Langbehn et al. \cite{langbehn2018evaluation}, Farmani et al. \cite{farmani2018discrete}, and Sarupuri et al.\cite{sarupuri2017trigger} concentrates efforts on strategies to resolve CS tied to locomotion in virtual reality environments.  Besides, Berthoz et al., \cite{berthoz1975perception} Pavard et al., \cite{pavard1977linear} and Bouyer et al. \cite{bouyer2017inducing} applied haptic feedback to cover this issue. Our proposed guideline (section \ref{lab:proposedGuideline}) detail these and other works considering their strategies.

\subsection{VR-based serious games}

In healthcare, serious games aimed at patient rehabilitation are top-rated for many reasons \cite{sandlund2009interactive}. In these rehabilitation activities, frequent patients give up long-term treatments (such as physiotherapy) because they feel tedious after many exercise sessions \cite{moya2011use}. 
The serious games can motivate patients during the rehabilitation tasks, mainly in long-term treatments where the patient needs to train their motor abilities \cite{burke2009optimising}. 

Some studies \cite{ferreira2020adaptive,liao2021virtual} designed virtual reality-based serious games to enhance the execution of these movements using virtual reality and HMDs. In Ferreira et al. \cite{ferreira2020adaptive} work, they developed a serious VR-based game for upper limb rehabilitation. In this case, the primary purpose of the developed games is to keep patients engaged without developing any frustration or boredom during the gameplay. They used an immersive Tetris-like game to execute movements that strengthen patients' muscles of the arm and shoulder, stimulate their cognitive activity, and stimulate the patient's sensory inputs. The 15 participants were exposed to a VR-based game for an average of 28 minutes and 46 seconds.

Liao et al. \cite{liao2021virtual} also proposed a serious VR-based game for upper limb therapy. The serious game developed is composed of three mini-games, where each one is designed for a specific rehabilitation task related to daily life or a joint event (i.e., fishing). They use a scoring system to reward patients after completing tasks in games. The authors did not mention the number of participants and the exposure time in their VR-based game.  

Pereira et al. \cite{pereira2021hand} developed the Stable-Hand VR, a serious game developed to help patients in hand rehabilitation processes. This system uses an HMD to render immersive content and a leap controller to track and manage the patient hand's movements. This serious VR-based game's tasks are generally aimed at pinching gestures. The experiment process in this work was made with 7 participants (in both works) in two 20 minutes sessions with one day break. 
According Pereira et al. \cite{pereira2021hand}, 3 participants mentioned discomfort tied to HMD use, specifically during the transition between scenes in the game menu.

Luo et al. \cite{luo2020vr} designed the VR driving serious game focused on rehabilitation of patients with Hemiplegia, a condition caused by brain harm that results in weakness, stiffness (spasticity), and lack of control in one side of the body. They use a VR serious game to create rehabilitation sessions associated with upper limp recovery treatment. In summary, the patient has the task of controlling a virtual car to collect coins and avoid virtual obstacles. The game has two different play modes, one with a fixed speed and the other with a controllable speed (similar to conventional race games). They produced two modes to adapt the application with different limb disorder levels. Also, the application has an HMD and 3D annular projection screen modes. According to the authors, compared to traditional rehabilitation activities, the VR game increased the interest and confidence of patients during the rehabilitation processes, but no CS mitigation technique was applied.


Although all the related works are focused on patient rehabilitation (and related to health), only the work by Pereira et al. \cite{pereira2021hand}  mentioned the concern with symptoms of the discomfort of their patients during the use of HMD device. In addition, none of the studies mentioned applied one or more strategies to avoid discomfort in their patients during HMDs use. Additionally, a recent study related an option to use the same developed VR-based game in a non-immersive way (with a standard screen) and the same controllers (VR motion controllers) to avoid CS \cite{lanzoni2022design}.

The studied serious games in this work do not endeavor to avoid CS using strategies. On the other hand, commercial games essentially use CS minimization techniques. We assume that VR-based serious games for rehabilitation and traditional VR games have similar gameplay, design, and development bases. For this reason, we propose a guideline framework with strategies to combat CS in VR-based games (including those for rehabilitation).

\section{Proposed Guideline Framework}
\label{lab:proposedGuideline}


As previously mentioned, many related works have investigated the influence of CS aspects in VR environments. However, in this study, we are interested in investigating how such aspects of CS have impacted the user experience with virtual reality serious games and how game designers could mitigate such impacts.


Following a previous review of the literature \cite{porcino2017minimizing,  porcino2021cybersickness}, we propose a guideline strategy that may be used as a framework for avoiding CS in VR applications and games.
In this section, we present our proposed guideline framework based on causes described in a previous work \cite{porcino2017minimizing,porcino2020using}. Based on the referenced serious games for health applications activities, we propose here that the causes of CS may be influenced by the following variables: locomotion, acceleration, the field of view, depth of field, degree of control, duration use time, latency-lag (including rendering optimization), static rest frame, and camera rotation. Our framework first suggests identifying when these situations are present in the application, marking them as necessarily desirable or optional, according to the serious game required activities. We also suggest listing these situations in a priority list to give more attention to those at the top of the list. Our framework suggests the designer consider the following guidelines for each of these variables.

\subsection{Locomotion}
\begin{figure}[ht]
\centering
\includegraphics[width=1\linewidth]{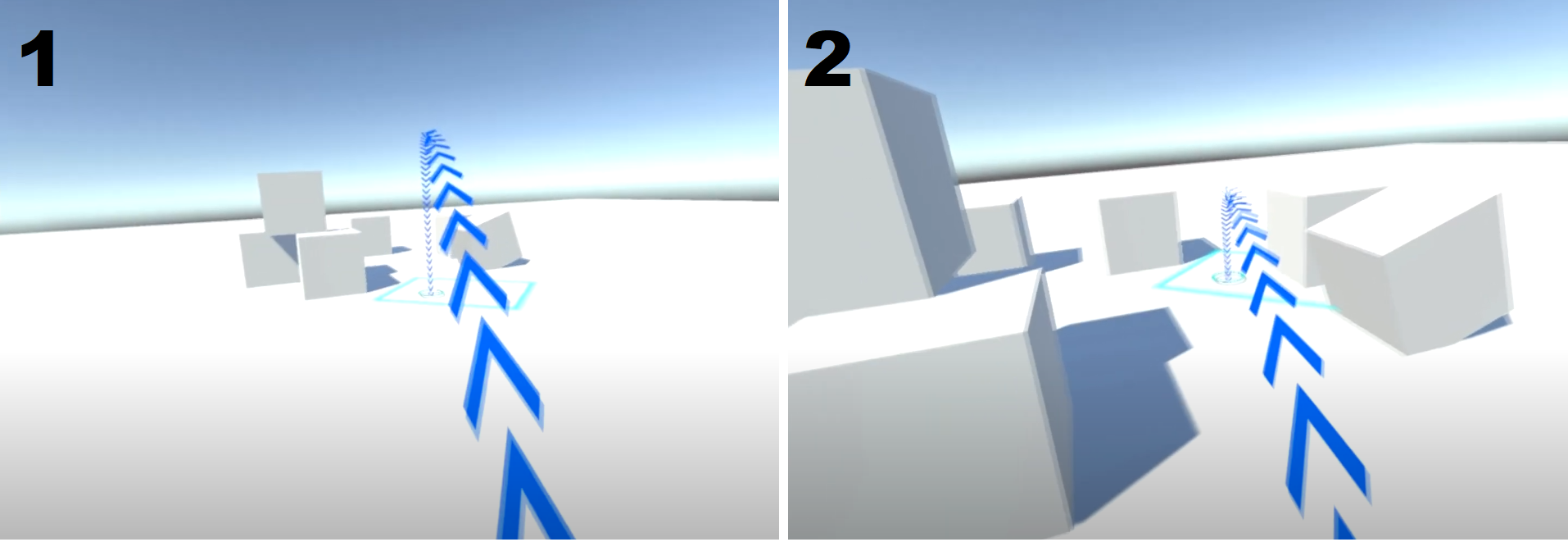}
\caption{Example of teleportation strategy. (1) The user point to an specific location in VR. (2) User are immediately transported to the aimed location.}
\label{fig:teleportation}
\end{figure}

We start our framework with the consideration of the user movement. In this sense, Teleportation strategies are one of the most effective strategies for avoiding CS related to translation, and our solution claims that developers should first consider if teleportation fits the application's requirements (Figure \ref{fig:teleportation}).

In teleportation, users can travel great distances by specifying the trip's destination point with the help of a marker \cite{langbehn2018evaluation}. This strategy works as follows: using a controller, the user points to the destination location and squeezes a trigger button, which immediately transports the user to the new location, also called "pointing and teleport". Another strategy called "trigger walk" uses the concept of natural walking to reach a destination. In this case, the user uses VR control triggers instead of legs to move around. Each user's hands handle each control in a relaxed and comfortable position (with minimum energy consumption). The user moves closer to the direction indicated at each pull of the trigger \cite{sarupuri2017trigger}. The VR-based serious games that able players to virtually walk need to attempt locomotion techniques to avoid CS.


\subsection{Acceleration}
\begin{figure}[ht]
\centering
\includegraphics[width=1\linewidth]{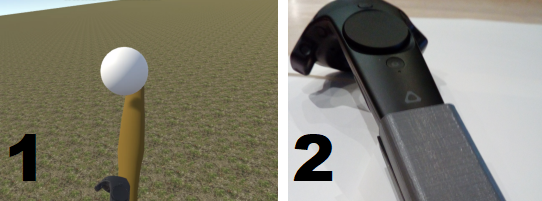}
\caption{The virtual reality game (1) and a (2) HTC Vive controller enhanced with haptic feedback actuators in Ryge et. al \cite{ryge2017effect}.}
\label{fig:hapticfeedback}
\end{figure}

Following locomotion, we suggest continuing the design process considering acceleration issues. According to \cite{berthoz1975perception}, 
it is possible to induce a sensation of movement using a visual response (haptic feedback example in Figure \ref{fig:hapticfeedback}) \cite{ryge2017effect}. 

According to \cite{pavard1977linear}, the human visual system can adapt to illusive motion but not acceleration. Various applications of VR (e.g., games and training applications) require support for haptic perception. This is because haptic perceptions can induce the sensation of acceleration in its users. When correctly applied, artificial acceleration sensation can help avoid sensory conflict.
In some virtual environments (racing games), it is possible to minimize CS problems using haptic responses. Haptics is a way of transmitting physical sensations to the user, which are compatible with those captured by the user’s visual system.
\cite{bouyer2017inducing} used haptic feedback outside a VR environment while still managing to provide users with an enhanced sense of reality. According to  \cite{plouzeau2018using}, measuring CS acceleration using electro-dermal activity (EDA) is possible. Plouzeau et al. changed and adjusted the acceleration to visualize EDA changes. When EDA values increase, the acceleration decreases proportionately. According to research \cite{tran2017subjective}, the more predictable the camera movement and acceleration, the lesser the CS effects will be. When accelerations situations are necessary for the required simulation, our framework suggests building or including proper interfaces, such as real ergonomic bicycles, adapted wheelchairs, or adapted mats.


\subsection{Field of view}
\begin{figure}[ht]
\centering
\includegraphics[width=1\linewidth]{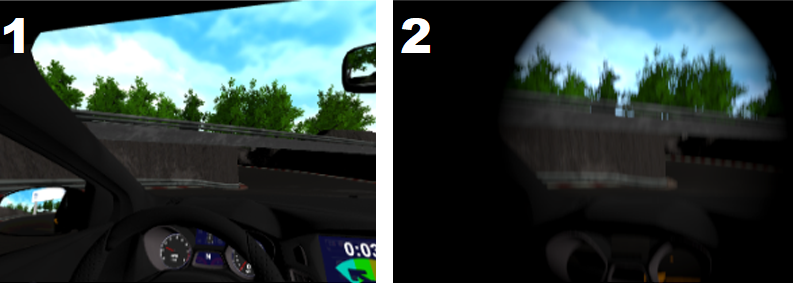}
\caption{(1) A VR race gameplay on a curved track with no appliance of guidelines. (2) The race gameplay using rotational blurring and tunneling, activated during the gameplay in  \cite{porcino2022identifying}.}
\label{fig:fov}
\end{figure}

Applying strategies that manipulate the field of view (FoV) in commercial games is quite common \cite{EagleFlight}. Vignette is a strategy to shorten the FoV gradually, thus reducing discomfort in VR environments  \cite{fernandes2016combating}. This strategy is a variation applied in  \cite{bolas2017dynamic}, where the size of the vignette and dynamic FoV is related to the camera acceleration values. Tunnel or Tunneling (illustrated in Figure \ref{fig:fov}) is also used to solve locomotion problems.

Such a strategy reduces the size of the user's FoV at the moment of the locomotion, thereby minimizing sensory conflict problems. Similar to the vignette, the tunnel significantly reduces the FoV. However, it is only applied during locomotion. Design projects should select all the translation stages and mark when such strategies could be used, considering the focus elements required for that action. Developers should avoid presenting new elements to the application during this moment. This strategy is required to improve locomotion techniques if the VR-based therapy players have to walk virtually. Additionally, VR-based therapies with camera movements can also benefit from this strategy.

\subsection{Depth of field}
\begin{figure}[ht]
\centering
\includegraphics[width=1\linewidth]{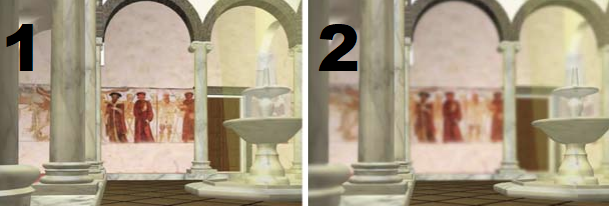}
\caption{(1) A VR application without depth of field effect. (2) The VR application using depth of field  focusing on the fountain \cite{zhou2007accurate}.}
\label{fig:dof}
\end{figure}

Some studies include a depth of field simulation (Figure \ref{fig:dof}) agent with image blur to minimize the convergence and accommodation problems \cite{carnegie2015reducing, porcino2016dynamic}. 

The solution presented by  \cite{carnegie2015reducing} pointed to decreased discomfort in HMD applications. Specifically, they suggested a GPU-based solution to simulate depth of field in these applications.
In our initial work, we developed a focus and region of interest dynamics model for the visualization of objects in VR \cite{porcino2016dynamic}. Unlike Carnegie, we used the term "dynamic" to suggest that the model moves the region of interest in the 3D scene using the application. This prototype simulates a visual focus self-extraction tool, limiting the region of interest in the visual field. The model uses the region of interest to determine the depth of field effects in real-time, minimizing discomfort when using HMD. Additionally, we designed a methodological guide for CS minimization on VR application \cite{porcino2017minimizing}.
On the other hand, field depth simulation strategies can produce low frame rates, inducing high latency, thereby causing CS. In the work of \cite{konrad2017accommodation} and \cite{padmanaban2017optimizing}, an approach was adopted to apply a simulated depth of field using an external interface, solving the problem of low frame rate. However, such a strategy is contingent on the application of specific hardware. A recent work reduced the spatial resolution in the HMD's peripheral regions (using foveated rendering) and applied the depth of field simulation effect  \cite{hussain2021mitigating}. Most observed VR-based serious games for rehabilitation render sharp images in HMDs, that trigger CS. In VR therapy, the lack of correct simulation of DoF can negatively affect the VR therapy, causing discomfort in the patient.

\subsection{Duration use time}
The study by  \cite{melo2018presence} relates the exposure time with discomfort manifestation and suggests short-term or interval virtual experiences. This principle suggests that VR applications could avoid CS symptoms if paused periodically. Consequently, the application should allow users to interrupt the experience to take a rest and then be able to return to the exact point paused before. According to \cite{stanney1997psychometrics}, CS can be minimized by exposing non-experienced users gradually to VR, which means a VR experienced user can have less susceptibility to CS than non-experienced users. Our framework suggests measuring the required time for the activities, and in cases when they take more than 10 minutes \cite{ramsey1999virtual}, we suggest to divide activities into time slots that may be interrupted without any immersive breakthrough.

\subsection{Degree of freedom (control)}
\begin{figure}[ht]
\centering
\includegraphics[width=1\linewidth]{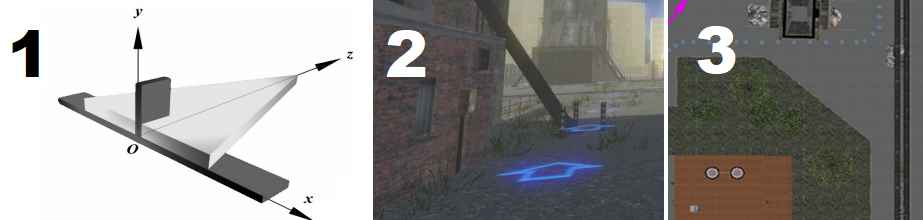}
\caption{(1) Design of the Virtual guiding 3D avatar by \cite{lin2004virtual}. (2) A view of  guiding avatar (3d arrows) use in first person perspective, and (3) a map of virtual environment showing multiple guiding arrows around the VR environment \cite{cao2018visually}.} 
\label{fig:guidingAvatar}
\end{figure}

Degrees of freedom (3-Dof or 6-Dof) is an essential concept in VR that allows natural human movement to be converted into simulated movement in the VR application \cite{nakka2018six}.
Low-cost HMDs (such as Google Cardboard, Oculus Go, and Samsung Gear VR) has 3 degrees of freedom tracking (3Dof). In other words, it is possible to look at 360 degrees, for example, looking left or right, rotating their head up or down, and pivoting left or right. Consequently, the user can feel disorientation by the limited movement. For 3-Dof HMDs, anticipating and preparing the user’s visual motion experience can reduce the lack of control by the user and, consequently, the discomfort. Having a virtual guiding avatar, which prepares the user’s visual motion experience in the virtual environment, is a result of VIMS reduction on Lin et al. \cite{lin2004virtual} (illustrated in Figure \ref{fig:guidingAvatar}).

\subsection{Latency-lag}
The asynchronous time warp is a method for overcoming HMD's latency \cite{kemeny2020getting}  by improving a rendered (warped) image based on the latest head-tracking data. According to \cite{van2016asynchronous} 
, this method is based on augmented reality "CamWarp" (that is applied in see-through augmented reality devices) and also reduces discomfort in VR environments. In other words, this strategy drives the user to see their movement in VR without diverging between the physical movement (made by the user's body) and the virtual movement (visualized by the user's eyes on the HMD display).



\begin{figure}[ht]
\centering
\includegraphics[width=1\linewidth]{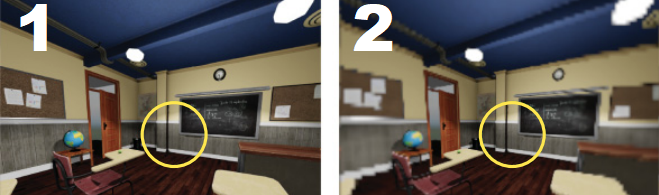}
\caption{Albert et al. \cite{albert2017latency} produced a scene with different foveation strategies. In (1) the original render without any effect, and in (2) they exaggerated the magnitude of foveation to illustrate the effect. The center of gaze (in yellow) is the region with high-resolution.}
\label{fig:foveated}
\end{figure}

Moreover, foveated rendering is a performance optimization based on the well-known degradation of peripheral visual acuity (illustrated in Figure \ref{fig:foveated}).

This strategy reduces the computational costs by rendering a high-resolution image in the user's center of attention (foveal region) and a low-resolution image in the periphery \cite{albert2017latency}. In VR, foveated rendering is an excellent graphics optimization that helps to reduce system latency and virtual reality application performance. 
Similar to "CamWarp" strategy, the "Time-Warped Foveated Rendering" \cite{franke2021time} strategy produces a visually plausible result if compared to regular rendering. In summary, Franke et al. \cite{franke2021time} produced a foveated rendered scene without being noticeable to the observer's eye with a speed-up of almost 1.6X than the original render.
Consequently, both strategies contribute to minimizing CS effects tied to latency. While VR-based serious games for rehabilitation are simpler than commercial entertainment games in the context of computational complexity (that affects the game performance), in VR-based therapy, the game performance is also critical because it directly affects the patient's health in case of triggering CS symptoms.

\subsection{Static rest frame}
\begin{figure}[ht]
\centering
\includegraphics[width=1\linewidth]{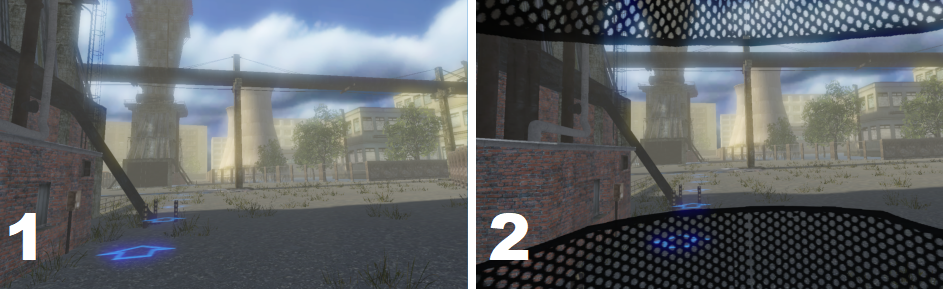}
\caption{(1) A VR application without static rest frame. (2) The VR application using  static layer as rest frame \cite{cao2018visually}.}
\label{fig:restframe}
\end{figure}

According to studies \cite{sharples2008virtual, kim2012comparison}, people show longer tolerance to discomfort during experiences based on VR projections (example: Cave Automatic Virtual Environment \cite{muhanna2015virtual}). One of the biggest differences between VR and projection-based systems is rest frames. In projection-based systems, the screen edges and visible real-world elements beyond the screens act as rest frames. This raises the hypothesis that the simulation of rest frames in virtual environments (Figure \ref{fig:restframe}) can create comfortable experiences \cite{lin2004virtual}.

However, adding elements to create a false rest frame that hides part of the screen may not be a good strategy for all VR applications. It can work well for racing games, where the player is naturally inserted into a car. However, this approach may not work well for games with first-person cameras, as they create strange circumstances for the player. Our framework suggests that the designer consider when this can be possible or not during the experience.

\subsection{Camera rotation}
\begin{figure}[ht]
\centering
\includegraphics[width=1\linewidth]{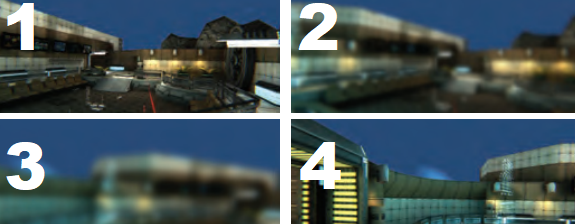}
\caption{From (1) to (4), a blurring effect application during a rotation movement in virtual reality environment \cite{budhiraja2017rotation}.}
\label{fig:blurring}
\end{figure}

Several other works applied various strategies such as head movement amplification, whereby individual movements are amplified in VR  \cite{kopper2011towards,plouzeau2018using}. Another example is the blurring rotation (Figure \ref{fig:blurring}), a strategy implemented by  \cite{budhiraja2017rotation} that uniformly applies Gaussian blurs based on the magnitude of acceleration and rotation values. There are also experiments deploying more basic strategies that lock the users’ heads to avoid rotational movements. 

According to  \cite{kemeny2017new}, such a strategy reduces the CS manifested during rotation by 30\% compared with the use of controls to perform rotational movements. Nevertheless, the authors concluded that participants found the strategy non-intuitive because it reduced the sensation of presence in the virtual environment. It is worth noting that both this strategy and rotation blurring only apply to rotational movements. We suggest using the blurring effect in a virtual reality experience during a rotation movement.

\section{Challenges to Minimize CS}

While many signs of progress related to immersion in VR had been achieved in computer graphics visualization, audio, smart devices, video tracking, and interfaces between many systems, there are still many remaining challenges to minimize CS in HMDs, primarily related to the early identification of situations that cause discomfort in the user. 
This section presents challenges such as CS prediction, data labeling for machine learning studies, the lack of underrepresented groups' data analysis, and the variability of tasks in virtual environments.


\begin{enumerate}
    \item \textbf{Cybersickness prediction} - Although deep learning classifiers are the most suitable approach for CS prediction, deep neural networks are black boxes that are very difficult to grasp. On the other hand,  symbolic machine learning algorithms enable a straight understanding of decision paths \cite{porcino2020using}. The human-readable characteristic of symbolic machine learning models such as decision tree and random forest can help understand the discomfort manifestation reasons \cite{porcino2021symbolic}. Additionally, a recent approach applies strategies to make deep learning models explainable \cite{xie2020explainable}, although the literature is still scarce on the topic. Explainable AI can be an excellent way to better the two worlds: the human understanding of symbolic machine learning with the accurate prediction of deep neural networks. 
    
    \item \textbf{Data labeling} - Another critical problem in CS research is associated with data labeling. Researchers generally collect verbal \cite{porcino2020using, kim2019deep}, haptic, or brain signal feedback to construct the ground truth of sickness. While verbal feedback is highly subjective and different for each participant, collecting haptic feedback when participants are in discomfort can often be corrupted by the delay associated with participant feedback. 

    \item \textbf{Data from underrepresented groups} - Another not resolved challenge is acquiring data considering underrepresented user-profiles \cite{peck2020mind} such as specific impairments or health issues \cite{lorenzini2021factors}. 

    \item \textbf{Tasks in VR} - A recent work \cite{grassini2020modern} pointed out that specific tasks can produce different results of CS for different user profiles and groups. Additionally, in a recent work \cite{porcino2022identifying} we suggested the importance of a better understanding of the correlation between profile attributes (such as gender, age, XR experience, and other individual characteristics) and gameplay elements, and also how outcomes obtained from profile attributes can be used to tag XR experiences according to distinct groups of users.     In this context, studies involving machine learning models combined with an evaluation of individual tasks associated with CS causes in XR games may produce a more profound study isolating any other XR possible influences on CS results. 
    
    
\end{enumerate}

\section{Discussion}
Although several strategies minimize cybersickness, it is still impossible to eliminate all symptoms from HMD users. This problem occurs because each user has a different level of susceptibility to discomfort, leading many researchers to look for solutions involving machine learning (deep neural networks), as seen in this work. 

The immersive serious games described in the related works did not use any strategy to minimize CS. This problem worsens when the serious game is associated with some recovery therapy. While the treatment in a virtual reality environment can be beneficial for the patient, it can generate feelings of discomfort in users if CS issue is not considered, as reported by Pereira et al. \cite{pereira2021hand}. For this reason, immersive serious games (especially those related to health) must follow guidelines to improve their users' experience in VR environments.

Moreover, we did not find any standard methodology for CS data recording or experiments conducted along with this study. Creating a standard methodology to collect CS data can be challenging because the same sickness level can differ among different participants. In other words, considering a scale from 0 to 3 (none, slight, moderate, and severe), one person can consider scale one as slight, but another, more susceptible, will fall as three or severe. 

Furthermore, collecting haptic feedback while participants are feeling discomfort, this data can be corrupted by delayed (i.e., CS cases delays in response time) or random responses. In other words, the moment the participant feels the pain may not be the exact moment the CS cause was triggered in the participant. Additionally, monitoring methodologies used to record the human brain's electrical activity (such as Electroencephalography) have more accurate data, as they capture more precise data of the user's physical and mental state, and many diseases and brain problems are diagnosed through the evaluation of such devices' data \cite{sanei2007eeg}. However, these medical tools are expensive and unavailable to the general public.

In summary, it is necessary to standardize the procedures, equipment, and materials to collect CS data and the methods used to classify it. 

\section{Conclusion and Future Work}
This work proposes an objective guideline process based on an extensive literature review to avoid CS using software and hardware approaches. Besides, CS can be triggered due to different factors \cite{porcino2021symbolic}. Eventually, a single
cause might influence more in triggering CS than others. For this reason, different combinations of strategies can be applied according to the game content or user susceptibility to CS. In other words, different VR-based serious games could lead to different strategy recommendations.

We covered the main strategies reported in the literature to minimize CS in VR compatible with immersive serious games. Additionally, we categorize the main challenges to minimizing CS into 4 different items, although others might be. Specifically, this classification is not exhaustive, and many other aspects could also be included in future works, such as human behaviors, hardware, interface, and other related issues. 

A straightforward way is to explore gender differences. A recent work  \cite{grassini2020modern} pointed out that specific tasks can produce different results of discomfort for different user profiles and user groups, regarding and not limited to: gender, age, or health issues. Our symbolic machine learning approach can also help to further analysis. 

For this reason, the next step of this research is to propose a virtual reality game that explores specific tasks individually tied to specific CS causes and strategies with a long exposure. We intend to conduct an evaluation study, considering individual tasks associated with CS causes and strategies. We believe it is possible to produce a more profound study isolating any other VR possible influences on CS results  \cite{porcino2020minimizing,porcino2021symbolic}. 


At this moment, we believe this work can contribute to 
designers and researchers to enable the right strategy to avoid CS in  VR serious games and other VR applications.

\bibliographystyle{abbrv}
\bibliography{references}

\end{document}